\documentclass[9.96pt, a4paper, twocolumn]{IEEEtran}

\usepackage{amsmath,epsfig,amssymb,verbatim,amsopn,cite,multirow}
\usepackage{amsthm}
\usepackage{balance}
\usepackage{multirow}
\usepackage[usenames,dvipsnames]{color}
\usepackage[all]{xy}  
\usepackage{url}
\usepackage{amsfonts}
\usepackage{amssymb}
\usepackage{epsfig}
\usepackage{epstopdf}
\usepackage{bm}
\usepackage{balance}
\usepackage{graphicx}
\usepackage{subcaption}
\usepackage{footnote}
\usepackage{cancel}
\usepackage{algorithm,algorithmic}

\usepackage[left=1.41cm,right=1.41cm,top=1.8cm,bottom=4.3cm]{geometry}

\newcounter{mytempeqcounter}

\newcommand{\qa}{{\bf a}}

\newcommand{\qe}{{\bf e}}

\newcommand{\qg}{{\bf g}}
\newcommand{\qh}{{\bf h}}

\newcommand{\qu}{{\bf u}}

\newcommand{\qw}{{\bf w}}
\newcommand{\qx}{{\bf x}}

\newcommand{\qH}{{\bf H}}
\newcommand{\qI}{{\bf I}}

\newcommand{\PMRT}{\mathtt{PMRT}}
\newcommand{\PZF}{\mathtt{PZF}}

\newcommand{\SINRE}{\mathrm{SINR}_{E}}
\newcommand{\SINRone}{\mathrm{SINR}_{1}}
\newcommand{\SINRk}{\mathrm{SINR}_{k}}

\newcommand{\Betalk}{\beta_{l,k}}
\newcommand{\betalE}{\beta_{l,E}}

\newcommand{\BphiE}{\boldsymbol{\phi}_E}
\newcommand{\Bphik}{\boldsymbol{\phi}_k}

\newcommand{\Bphione}{\boldsymbol{\phi}_1}
\newcommand{\Bvarepsilon}{\boldsymbol{\pi}}

\newcommand{\betalk}{\beta_{l,k}}

\newcommand{\betalone}{\beta_{l,1}}
\newcommand{\hlmk}{\qh_{l,k}}
\newcommand{\glmk}{\qg_{l,k}}
\newcommand{\hlmE}{\qh_{l,E}}
\newcommand{\glmE}{\qg_{l,E}}

\newcommand{\hhatlmk}{\hat {\qh}_{l,k}}

\newcommand{\hhatlm}{\hat{\qh}_{l, 1}}
\newcommand{\hhatlmE}{\hat{\qh}_{l, E}}

\newcommand{\gamalmk}{\gamma_{l, k}}
\newcommand{\gamalmE}{\gamma_{l, E}}

\newcommand{\thetpk}{\theta_{p, k}}
\newcommand{\thetlk}{\theta_{l, k}}
\newcommand{\thetpt}{\theta_{p, t}}
\newcommand{\thetlt}{\theta_{l, t}}

\newcommand{\tausl}{|\mathcal{S}_{l}|}

\newcommand{\Sl}{\mathcal{S}_{l}}

\newcommand{\Ex}{\mathbb{E}}

\newcommand{\trace}{\mathrm{tr}}

\title{Joint Power Optimization and AP Selection for Secure Cell-Free Massive MIMO}

\author{Yasseen Sadoon Atiya, Zahra Mobini, Hien Quoc Ngo, and  Michail Matthaiou\\
	\small{
		Centre for Wireless Innovation (CWI), Queen's University Belfast, U.K.\\
		Email:\{yhimiari01, zahra.mobini, hien.ngo, m.matthaiou\}@qub.ac.uk
}}\normalsize
\allowdisplaybreaks

\begin{document}
\bstctlcite{IEEEexample:BSTcontrol}
\maketitle
\begin{abstract}
In this paper, we investigate joint power control and access point (AP) selection  scheme in a cell-free massive multiple-input multiple-output (CF-mMIMO) system under an active eavesdropping attack, where an eavesdropper tries to overhear the signal sent to one of the legitimate users by contaminating the uplink channel estimation. We formulate a joint optimization problem to minimize the eavesdropping spectral efficiency (SE) while guaranteeing a given SE requirement at legitimate users. The challenging
formulated problem is converted into a more tractable form
and an efficient low-complexity accelerated projected gradient (APG)-based approach is proposed to solve it. Our findings reveal that 
the proposed joint optimization approach significantly outperforms the heuristic approaches in
terms of secrecy SE (SSE). For instance, the $50\%$ likely SSE performance of the proposed approach is $265\%$ higher than that of equal power allocation and random AP selection scheme. 

\let\thefootnote\relax\footnotetext{The authors  are with the Centre for Wireless Innovation (CWI), Queen's University Belfast, BT3 9DT Belfast, U.K. email:\{yhimiari01, z.mobini, hien.ngo, m.matthaiou\}@qub.ac.uk.

This work is a contribution by Project REASON, a UK Government funded project under the Future Open Networks Research Challenge (FONRC) sponsored by the Department of Science Innovation and Technology (DSIT).
The work of Z. Mobini and H. Q. Ngo was supported by the U.K. Research and Innovation Future Leaders Fellowships under Grant MR/X010635/1. The work of M. Matthaiou has received funding from the European Research Council (ERC) under the European Union’s Horizon 2020 research and innovation programme (grant agreement No. 101001331).} 
\end{abstract}

\vspace{-1em}
\section{Introduction} 
CF-mMIMO has emerged as a promising wireless access technology for the next generations of wireless networks. In CF-mMIMO systems, a large number of APs are deployed over a certain area to communicate with a much smaller number of users in a user-centric scenario. Consequently, it can provide high coverage, spectral, and energy efficiency achieved by low-complexity signal processing~\cite{Hien:JWCOM:2017}. CF-mMIMO leverages its architecture to provide efficient communication for all users in the network. However, this makes it more susceptible to eavesdropper (Eve) attacks due to the short distance between the APs and users. Therefore, the secrecy aspect of CF-mMIMO has attracted researchers’ attention in recent years. 

There are a few works in the context of secure CF-mMIMO systems. Specifically, in~\cite{Timilsina:GC:2018}, a secrecy performance comparison between CF-mMIMO and co-located massive MIMO systems was performed by deriving the SSE expression under active pilot attacks. Then, using the same setup of~\cite{Timilsina:GC:2018}, Hoang \emph{et al.} in~\cite{Hoang:TCOM:2018}, proposed an approach for detecting the presence of the active Eve and designed  
power control schemes with different objectives including maximizing the achievable SE of the legitimate user who is under attack, maximizing the SSE, and minimizing the total power at all APs subject to the constraints on the SE of the users and Eve. Later, the authors in~\cite{Salah:Globecom:2020} investigated the secrecy performance  of CF-mMIMO utilizing non-orthogonal pilot sequences in the uplink training phase and proposed an approach for pilot transmission in the downlink phase. In addition, the authors in~\cite{Alageli:TIFS:2020}  addressed the problem of joint downlink power control and data transfer of CF-mMIMO under active eavesdropping. The effect of hardware impairments on the secrecy performance of a CF-mMIMO system was studied in~\cite{Zhang:SYS.2020}. Moreover,~\cite{Zhang:TVT:2010} studied the effect of radio frequency (RF) impairments and low-resolution analog-to-digital converters  on the secrecy performance of CF-mMIMO system experiencing active eavesdropping.
However, all the aforementioned works focused on power control designs, assuming that all APs serve all users.

Numerous AP selection schemes have been introduced in the literature~\cite{Hien:Asilomar:2018,Ammar:GLOBSIP:2019,Buzzi:TWC:2020,Chen:JSAC:2021,Hao:2023:EUSIPCO} with the objective of increasing the scalability of CF-mMIMO, while simultaneously improving its SE and energy efficiency. Nevertheless, a popular assumption in the existing literature   on secure CF-mMIMO systems is that  each user is served by all APs in the network. In practice, APs are located at a considerable distance from users, making minimal contributions to the SE, whereas  others are close to a potential Eve. Therefore, the efficient selection of APs to serve users, especially users under an eavesdropping attack, is a meaningful and important challenge. In line with this idea, in our very recent work~\cite{Yasseen:TWC:2023}, we proposed  a heuristic AP selection approach to improve the SSE and showed that AP selection can provide a noticeable secrecy improvement.



In pursuit of harnessing the inherent benefits of \emph{concurrently optimizing power allocation and AP selection}, in this paper we propose a  joint optimization approach based on the APG method to minimize the overheard signal-to-interference-plus-noise ratio (SINR) at Eve while ensuring a predetermined quality of service (QoS) constraint at legitimate users. The main contributions of this paper can be summarized as follows:
\begin {itemize}
\item We propose a joint  optimization approach for AP selection  and power optimization. The formulated problem is under a transmit power constraint and  individual QoS requirements of all legitimate users. 
\item  An efficient algorithm is then proposed to solve the challenging formulated mixed-integer non-convex problem.
\item Numerical results show that our joint optimization
approach remarkably improves the secrecy  performance of
the CF-mMIMO system and  outperforms the heuristic approaches. 
\end {itemize}

\textit{Notation:} Matrices are denoted by bold upper case letters, while bold lower case letters indicate vectors. The superscript $(\cdot)^\text{H}$ stands for the conjugate-transpose (Hermitian); we use $\mathbb{C}^{L\times N}$ to denote a $L\times N$ matrix; an$M\times M$ identity matrix is represented by $\qI_M$; $\trace(\cdot)$ refers to the trace operation. The statistical expectation is denoted by $\Ex\{\cdot\}$. Finally, a zero mean circular symmetric complex Gaussian distribution with variance $\sigma^2$ is denoted by $\mathcal{CN}(0,\sigma^2)$.	

\section{System Model}~\label{sec:sysmodel}
A CF-mMIMO system with $L$ APs, $K$ single-antenna  users and time division duplex (TDD) mode is considered. Each AP $l$, $l \in \mathcal{L}=\{1,\ldots, L\}$, is equipped with $M$ antennas such that
$LM >> K$. There is also an active single-antenna Eve  which is attempting to intercept the data destined to a specific user. Without loss of generality, let us assume that this is user ${1}$. The $M\times 1$ channel vector from the $l$-th AP to the $k$-th user, $k \in \mathcal{K}=\{1,\ldots,K\}$ and to Eve are  given by
\begin{align}~\label{eq:hlmk}
		\hlmk=\sqrt{\betalk} \glmk
~\quad~\text{and}\quad  \hlmE=\sqrt{\betalE} \glmE,
\end{align}
respectively, where $\betalk$ and $\betalE$  denote the corresponding large-scale fading coefficients, while $\glmk\sim \mathcal{CN}(0,\qI_M)$ and $\glmE\sim \mathcal{CN}(0,\qI_M)$ are the complex Gaussian random vectors with covariance matrix $\qI_M$ which represent small-scale fading.
We consider a CF-mMIMO system with an AP selection scheme in which the
serving APs of each user are selected from a set of candidate APs to maximize the secrecy rate according to the system QoS requirement. In particular, we define the binary variables to indicate the AP-user association as
\begin{align}
\label{a}
a_{l,k} \triangleq
\begin{cases}
  1, & \text{if  user $k$ is served by $l$-th AP,}\\
  0, & \mbox{othewise}.
\end{cases} 
\end{align}
The considered TDD transmission includes three main phases: uplink training phase for estimating channels, downlink data transmission, and uplink data transmission. Here, we focus on the downlink data transmission, and hence, neglect the uplink data transmission.
\vspace{-0.35em}
\subsection{Uplink Training and Downlink Data Transmission}
In the uplink training phase, we assume that all users transmit their pairwisely orthogonal pilot sequences of length $\tau_p\geq K$ to all the APs in each coherence block. Denote the pilot sequence sent by the $k$-th user by $\Bphik\in\mathbb{C}^{\tau_p \times 1}$. Note that as discussed in\cite{Hoang:TCOM:2018}, all pilot sequences are publicly known, and hence, Eve can transmit the same pilot as the targeted user ${1}$ by setting $\BphiE = \Bphione$ to contaminate the channel estimates between all APs and user $1$. Subsequently, Eve overhears the signals sent to user ${1}$.  By utilizing the minimum-mean-square-error
(MMSE) method   at the $l$-th AP to estimate the channel between the $l$-th AP and $k$-th user, $\hhatlmk$, we have $\hhatlmk\sim\mathcal{CN}(0,\gamalmk\qI_M)$, where
\vspace{-0.35em}
\begin{align}~\label{eq:gamalmk}  
		\gamalmk\triangleq 
  \begin{cases}
  \frac{\tau_{p} \rho_{u} \betalk^{2}}{\tau_{p} \rho_{u} \betalk+1}, & k \neq 1, 
  \\ 
  \frac{\tau_{p} \rho_{u} \betalone^{2}}{\tau_{p} \rho_{u} \betalone+\tau_{p} \rho_{E} \betalE+1}, 
  & k=1,
		\end{cases} 
\end{align}
where $\rho_{u} \!\triangleq\! P_{u} / N_{0}$ and $\rho_{E} \!\triangleq\! P_{E} / N_{0}$, while  $P_u$ and $P_E$ are respectively the
transmit powers of users and Eve, and $N_0$ is the average noise power at the APs. Let $\alpha_{l,1}=\big(\rho_{E} \betalE^{2}\big) /\big(\rho_{u} \betalone^{2}\big)$, then the estimated channel and the mean-square of the estimates between the $l$-th AP and Eve are respectively given by  
\vspace{-0.35em}
\begin{align}~
\label{eq:hhatlmE} 
		\hhatlmE&=\sqrt{\alpha_{l, 1}} \hhatlm,
  \\
\gamalmE&=\alpha_{l,1} \gamma_{l,1}.\label{eq:gammalmE} 
\end{align}

In the downlink data transmission phase, all APs transmit their signals to all users by exploiting the channel estimates obtained during the previous phase. Let $s_{k}$, where $\Ex\{|s_{k}|^{2}\}=1$, denote the data symbol sent to the $k$-th user, the precoded data signal sent by  the $l$-th AP to all users can be written as
\begin{align}~\label{eq:x_l} 
		\mathbf{x}_{l}=  \sum_{k\in \mathcal{K}} \sqrt{\rho_{d}}{\theta _{l,k}} \qw_{l, k} s_{k},
\end{align}
where $\rho_{d}$ is the maximum normalized transmit power of each AP, $\theta_{l,k}$ is the power control coefficient, and $\qw_{l, k} \in \mathbb{C}^{M \times 1}$, where $\Ex\{\|\qw_{l, k}\|^{2}\}=1$, is the precoding vector associated with user $k$. Here, we enforce 
\begin{align}\label{eq:assos}
    \theta_{l,k} = 0 , \mathrm{if}~a_{l,k}= 0~\forall l, k,
\end{align}
to ensure that if AP $l$ does not serve user $k$, the transmit power of AP $l$ to user $k$ is zero. Note that AP $l$ is required to satisfy the average normalized power
constraint, which can  be written
as the following per-AP power constraint 
\begin{align}\label{eq:Pconst}
\sum_{k \in \mathcal{K}} \theta_{l,k}^2 \leq 1, \forall l.
\end{align}
Hence, the signals received at the $k$-th user and Eve are, respectively, given by
\vspace{-0.25em}
	\begin{align}~\label{eq:z_k}   
		{z}_{k}&= \sqrt{\rho_{d}}\sum\limits_{l\in \mathcal{L}}\mathbf h^{\mathrm{H}}_{l,k}\mathbf{x}_{l}+{n}_{k},
  \\
		{z}_{E}&= \sqrt{\rho_{d}}\sum\limits_{l\in \mathcal{L}}\mathbf h^{\mathrm{H}}_{l,E}\mathbf{x}_{l}+{n}_{E},~\label{eq:z_E} 
	\end{align}

where $n_{k} \sim \mathcal{CN}(0,1)$ and $n_{E} \sim \mathcal{C N}(0,1)$. 


Here, we employ the local protective partial zero-forcing (PPZF) precoding scheme due to its excellent balance between performance and complexity \cite{Emil:TWC:2020}. With PPZF, each AP mitigates the interference that causes to the subset of users, while tolerates the interference that creates to other users. For each AP $l$, users are split, based on the channel gains $\Betalk$, $l \in \mathcal{L}, k \in \mathcal{K}$, into two groups: strong group $\Sl$ and weak group $\mathcal{W}_{l}$ with  $\Sl \cap \mathcal{W}_{l}=\varnothing$, and $\vert\Sl\vert+\vert\mathcal{W}_{l}\vert=K$. 
After grouping users, each AP utilizes partial ZF (PZF) and protective MRT (PMRT) schemes for precoding the signals sent to users in $\Sl$ and $\mathcal{W}_{l}$, respectively.  We note that, to implement PZF,  the number of antenna elements must meet the requirement  $M \geq |\Sl|+1$. Let the collective channel estimation matrix from AP $l$ to all users be $\hat\qH_l = [\hat\qh_{l,1}, \ldots, \hat\qh_{l,K}] \in\mathbb{C}^{ M \times K}$ where $\mathbf{e}_i$ represents  the $i$-th column of $\mathbf{I}_{K}$, and hence $\mathbf{E}_{\Sl}=	\big[\qe_i: i \in \Sl\big] \in \mathbb{C}^{ K \times |\Sl|}$. 
Also, let $\Bvarepsilon_{k}$ be the $j$-th column of $\mathbf{I}_{|\Sl|}$ and user $k$ corresponds to the $j$-th element of set $\Sl$, $j \in \{1, \ldots , |\Sl|\}$. Then, the precoding vector in \eqref{eq:x_l} is given by
\begin{align}~\label{eq:wlk_PPZF1}  
		\qw_{l, k}= 
  \begin{cases}
  \mathbf {w}^ {\PZF} _{l, k}, & \text{if} ~ k \in \mathcal {S} _{l}, 
  \\ 
  \mathbf {w}^ {\PMRT} _{l,k}, 
  & \text{if} ~ k \in \mathcal {W} _{l},
		\end{cases} 
\end{align}
\vspace{-0.1em}
where $\mathbf {w}^ {\PZF} _{l, k}$ and $\mathbf{w}^ {\PMRT} _{l,k}$, are respectively, given by
\vspace{-0.2em}
\begin{align}~\label{eq:wPZF}
\qw_{l, {k}}^{\PZF}&= \!\!\sqrt {\! (\!M\!\!-\!\! \tausl)\gamma_{l,j}} \hat{\qH}_{l} \mathbf{E}_{\mathcal{S}_{l}}\left(\mathbf{E}_{\mathcal{S}_{l}}^{\mathrm{H}} \hat{\qH}_{l}^{\mathrm{H}} \hat{\qH}_{l} \mathbf{E}_{\mathcal{S}_{l}}\right)^{-1} \Bvarepsilon_k,
\\
\mathbf {w}^ {\PMRT} _{l,j} &=\frac { \mathbf {B}_{l}\hat \qH_{l} \mathbf {e}_{j} } {\sqrt { (M- \tausl)\gamma_{l,j} }},\label{eq:wPMRT}
 \end{align}
with
	\begin{align}~\label{eq:Bl}
		\mathbf{B}_{l}=\mathbf{I}_{M}-\hat{\qH}_{l} \mathbf{E}_{\mathcal{S}_{l}}\left(\mathbf{E}_{\mathcal{S}_{l}}^{\mathrm{H}} \hat{\qH}_{l}^{\mathrm{H}} \hat{\qH}_{l} \mathbf{E}_{\mathcal{S}_{l}}\right)^{-1} \mathbf{E}_{\mathcal{S}_{l}}^{\mathrm{H}} \hat{\qH}_{l}^{\mathrm{H}}. 
	\end{align}
Accordingly, \eqref{eq:x_l} can be expressed as
\begin{align}~\label{eq:xl}
\mathbf {x}_{l} = \sum _{k \in \mathcal {S} _{l}}  \sqrt{\rho_{d}}{\theta_{l,k}} \mathbf {w}^ {\PZF} _{l, k} s_{k} + \sum _{j \in \mathcal {W} _{l}} \sqrt{\rho_{d}}{\theta _{l,j}} \mathbf {w}^ {\PMRT} _{l,j} s_{j}.
\end{align}
\vspace{-0.9em}
\subsection{Achievable SE at the Users and Eve}
Let us denote the set of AP indices that utilize PZF and PMRT in transmission by $\mathcal{Z}_{k}$ and $\mathcal{M}_{k}$, respectively,
\begin{align}\label{eq:Z_k}
&\mathcal{Z}_{k} \triangleq\left\{l: k \in \mathcal{S}_{l}, l=1, \ldots, L\right\}, 
\\
&\mathcal{M}_{k} \triangleq\left\{l: k \in \mathcal{W}_{l}, l=1, \ldots, L\right\},\label{eq:M_k}
\end{align}
with $\mathcal{Z}_{k} \cap \mathcal{M}_{k}=\varnothing$, and $\left|\mathcal{Z}_{k}\right|+\left|\mathcal{M}_{k}\right|=L$. 

Accordingly, the received signal at the $k$-th user in~\eqref{eq:z_k} can be reformulated as
\vspace{-0.15em}
\begin{align}
&z_{k}=\Bigg(\sum_{l \in \mathcal{Z}_{k}} \sqrt{\rho_{d}}{\thetlk} \qh_{l, k}^{\mathrm{H}} \qw_{l, {k}}^{\PZF}\notag+\sum_{p \in \mathcal{M}_{k}} \sqrt{\rho_{d}}{\thetpk} \qh_{p, k}^{\mathrm{H}} \qw_{p, {k}}^{\PMRT}\Bigg) s_{k} \\
&+\!\sum\limits_{t\in \mathcal{K} \atop t\neq k}
\!\Bigg(\!\sum_{l \in \mathcal{Z}_{t}} \!\sqrt{\rho_{d}}{\thetlt}  \qh_{l, k}^{\mathrm{H}} \qw_{l, {t}}^{\PZF}\notag\!+\!\!\!\!\sum_{p \in \mathcal{M}_{t}}\!\!\!\sqrt{\rho_{d}}{\thetpt} \qh_{p, k}^{\mathrm{H}} \qw_{p, {t}}^{\PMRT}\!\Bigg) s_{t} \!+\!n_{k}.
\end{align}
Using the use-and-then-forget bounding method~\cite{Hien:JWCOM:2017}, an achievable SE of the $k$-th user is expressed as $R_k= \log_2\left(1+\SINRk \right)$, where
\vspace{-0.45em}
\begin{align} ~\label{eq:SINR_k2} 
\SINRk= 
\frac
{\Big(\sum\limits_{l\in \mathcal{L}} \sqrt{\rho_{d}\left(M-\tausl\right) \gamma_{l,k}}\thetlk \Big)^{2}}
{\sum\limits_{t\in \mathcal{K}} \sum\limits_{l\in \mathcal{L}}\rho_{d}\thetlt^2\left(\beta_{l, k}-\delta_{l,k}\gamma_{l, k}\right) +1},
\end{align}
with $\delta_{l, k} \triangleq 1$ if  $k\in \mathcal{S}_{l}$ and $\delta_{l, k} \triangleq 0$ if  $k\in \mathcal{W}_{l}$~\cite{YASSEEN:2023:VTC}.

In addition, the received signal at Eve in~\eqref{eq:z_E} can be represented in the form of
\begin{align} ~\label{z_E2} 
z_{E} \!=\!& \sum\limits_{l\in \mathcal{L} } 
\!\sqrt{\rho_{d}}\theta_{l,1}\mathbf h_{l,E}^{\text{H}}  \qw_{l,1} s_{1}\! + \!\! {\sum\limits_{t\in \mathcal{K}  \atop t\neq 1}} 
\sum\limits_{l\in \mathcal{L} } 
\!\sqrt{\rho_{d}}{\thetlt}  \mathbf h_{l,E}^{\mathrm{H}}  \qw_{l,t}s_{t}\! +\! n_{E}.
\end{align}
Following~\cite{YASSEEN:2023:VTC},
the SE of Eve can be written as $R_E = \log_2\left(1+\SINRE \right)$, where
\vspace{0.15em}
\begin{align}~\label{eq:SINR_E2} 
\SINRE\!\!=\!\!\frac {\!\!\!\left(\sum\limits_{l\in \mathcal{L} } \!\!\sqrt {\!\rho_{d}(M\!\!-\!\!\tausl)\gamma_{l,E}}{\theta _{l,1}}\!\!\right)^{2}\!\!\!\!\!+\!\!\!\sum\limits_{l\in \mathcal{L}} \!\!\rho_{d}\theta_{l, 1}^2(\beta_{l,E}\!-\!\delta_{l,1}\!\gamma_{l,E}\!)}{{\sum\limits_{t\in \mathcal{K} \atop t\neq 1}} \sum\limits_{l\in \mathcal{L} }\rho_{d}\thetlt^2(\beta_{l,E}\!-\!\delta_{l,1}\gamma_{l,E})\!+\!1}. 
\end{align}
Now, the SSE  can be calculated as
\begin{align}~\label{eq:R_sec} 
\mathrm{SSE} = [R_1-R_E]^+  =  \left[\log_{2}\left(\frac{1+\SINRone}{1+ \SINRE} \right)\right]^+,
\end{align}
where  $[x]^+ = \max \{0, x\}$.

\section{Joint Power Allocation and AP Selection}
We now seek to jointly optimize the power control coefficients $\boldsymbol{\theta} \triangleq \{\theta_{l,k}\}$ and AP selection $\qa \triangleq \{a_{l,k} \}$ to minimize the achievable SE at Eve under a  total transmit power constraint  at each AP and a QoS requirement $\mathrm{R}_{\mathrm{th},k}$  at  each user $k$. The corresponding optimization problem becomes
\begin{subequations}\label{eq:minRE}	
\begin{align}
&\underset{\qa,\boldsymbol{\theta}}{\mathrm{min}}\quad
\mathrm{R}_E(\boldsymbol{\theta}),\\
 \label{eq:st0} \text {s.t.}~~ \quad 
  &\thetlk \geq 0, \forall l, k, \\
  \label{eq:st1}\hspace{3em}&\sum_{k \in \mathcal{K}} \theta_{l,k}^2 \leq 1, \forall l,
\\
     \label{eq:st3} \hspace{3em} &\mathrm{R}_k(\boldsymbol{\theta})  \geq\mathrm{R}_{\mathrm{th},k},~~ \quad \forall k,
    \\
\label{eq:st5} \hspace{3em}    &\left(\theta_{l,k}^2=0,
\forall k, \text { if } a_{l, k}=0\right), \quad \forall l,
         \\ 
\label{eq:st6}\hspace{3em} &\sum_{l \in \mathcal{L}} a_{l, k} \geq 1, \forall k,
\end{align}
\end{subequations}
 where the constraint in~\eqref{eq:st6} guarantees that each user is served by at least one AP. Problem~\eqref{eq:minRE}  is a nonconvex mixed-integer optimization problem and is difficult to solve. To deal with the the binary constraint in $\eqref{a}$,  we observe that $x \in$ $\{0,1\} \Leftrightarrow x \in[0,1]\quad \& \quad x-x^2 \leq 0$~\cite{Vu:IOT:2022},  and hence, we  replace $\eqref{a}$ with
\begin{align}\label{eq:Q(a1)}
& Q(\mathbf{a}) \triangleq \sum_{k \in \mathcal{K}} \sum_{l \in \mathcal{L}}\left(a_{l,k}-a_{l,k}^2\right) \leq 0, \\
\label{eq:Q(a2)}& 0 \leq a_{l,k}, \forall l, k, \\
\label{eq:Q(a3)}& a_{l,k} \leq 1, \forall l, k.
\end{align}
Based on \eqref{eq:st1}, \eqref{eq:st5} can be replaced by
\begin{align}\label{eq:Q(a4)}
\theta_{l,k}^2 \leq a_{l,k}, \forall l, k \text {. }
\end{align}
To this end, problem \eqref{eq:minRE} is equivalent to
\begin{align}\label{eq:minxf}
\min _{\mathbf{x} \in {\mathcal{C}}} \mathrm{R}_E(\boldsymbol{\theta}),
\end{align}
where $\mathbf{x} \triangleq\{\mathbf{a}, \boldsymbol{\theta}\},{\mathcal{C}} \triangleq\{\eqref{eq:st0},\eqref{eq:st1},\eqref{eq:st3},\eqref{eq:st6},\eqref{eq:Q(a1)}-\eqref{eq:Q(a4)}\}$ is a feasible set. 
To solve problem \eqref{eq:minxf}, we propose an approach based on the APG algorithm. The APG-based method has shown to be highly effective for resource allocation in cell-free massive MIMO, as it offers very good performance, and  provides much lower complexity than conventional successive convex approximation  (SCA) algorithms, especially when the network size is large \cite{Mai:TWC:2022,farooq2021utility}. 
\vspace{-0.65em}
\subsection{APG Method}
To begin, we modify the optimization problem~\eqref{eq:minxf} through a change in variables, enabling a more efficient computation of the function's gradient and the subsequent projection.
To this end, we first introduce a new variable
$z_{l,k}^2 \!\!\!\triangleq\!\! a_{l,k}, \forall l,k$, where
\begin{align}\label{eq:zlk}
0\leq z_{l,k} \leq 1, 
\end{align}
and define new notations as follows
\begin{itemize}
    \item \!\!$U_k(\boldsymbol\theta)$=$(\sum\limits_{l\in \mathcal{L}}\!\! \sqrt {\rho_{d}(\!M\!\!-\!\!\tausl)\gamma_{l,k}}{\theta _{l,k}})^2,$
\item \!\!$V_k(\boldsymbol\theta)$=$\sum\limits_{t\in \mathcal{K}}\sum\limits_{l\in \mathcal{L}} \rho_{d}\theta_{l, t}^2(\beta_{l,k}-\delta_{l,k}\gamma_{l,k})+1$,
\item \!\!$U_E(\boldsymbol\theta)$=$(\!\sum\limits_{l\in \mathcal{L}}\!\!\sqrt {\!\rho_{d}(\!M\!\!-\!\!\tausl\!)\!\gamma_{l,1}}\!{\theta _{l,1}}\!^2+\!\!\sum\limits_{l\in \mathcal{L}}\! \rho_{d}\theta_{l, 1}^2(\beta_{l,E}\!\!-\!\delta_{l,1}\gamma_{l,E})$, \item \!\!$V_E(\boldsymbol\theta)$=$\sum\limits_{t\in \mathcal{K} \atop t\neq 1}\sum\limits_{l\in \mathcal{L}} \rho_{d}\theta_{l, t}^2(\beta_{l,E}-\delta_{l,1}\gamma_{l,E})+1$.
\end{itemize}
\vspace{-0.25em}
Accordingly, the objective function in Problem~\eqref{eq:minxf} can be rewritten as ${R}_{E}(\theta)= \log_{2}\left(1+ \frac{U_E(\theta)}{V_E(\theta)}\right)$ and the corresponding constraint~\eqref{eq:st3} can be replaced by 
${R}_{k}(\theta)= \log_{2}\left(1+ \frac{U_k(\theta)}{V_k(\theta)}\right)>=R_{\mathrm{th}}$.
Now, a key challenge in developing an efficient APG 
algorithm for solving Problem~\eqref{eq:minxf} is the QoS constraint \eqref{eq:st3} as well as the AP association constraints \eqref{eq:st6}, \eqref{eq:Q(a1)}, and \eqref{eq:Q(a4)}. One possible way to address this problem is to incorporate these constraints into the objective function by introducing a penalty parameter, resulting in the formulation of the penalized problem~\cite{Mai:TWC:2022}.  The APG technique is subsequently utilized to address the penalized problem, while this process is iterated until a predefined stopping criterion is met. To this end,  for each constraint we introduce a  quadratic loss function as
\vspace{-0.1em}
\begin{itemize} 
 \item   $\Psi_{1}\!(\boldsymbol{\theta}) \!\triangleq\!\!\! \sum\limits_{k \in \mathcal{K}}\!\!\left[\max \left(0,\mathrm{R}_{\mathrm{th},k}\!-\!\mathrm{R}_k(\boldsymbol{\theta})\right)\right]^2$ for constraint \eqref{eq:st3}.
 
 \item $\Psi_{2}(\mathbf{z}) \triangleq \sum\limits_{k \in \mathcal{K}} \sum\limits_{l \in \mathcal{L}}\left(z_{l,k}^2-z_{l,k}^4\right)$, where $\mathbf{z} \!\triangleq\!\left[\mathbf{z}_1^T, \ldots, \mathbf{z}_L^T\right]^T$ with
    $\mathbf{z}_l \triangleq\left[z_{l,1}, \ldots, z_{l,K}\right]^T$ for constraint \eqref{eq:Q(a1)}. 

  \item $\Psi_{3}(\boldsymbol{\theta}, \mathbf{z}) \!\triangleq\! \sum\limits_{k \in \mathcal{K}}([\max(0,1-\sum\limits_{l \in \mathcal{L}} z_{l,k}^2)]^2+\sum\limits_{l \in \mathcal{L}}[\max (0, \theta_{l,k}^2\!-\!z_{l,k}^2]^2)$ for constraints \eqref{eq:st6} and \eqref{eq:Q(a4)}.
\end{itemize}
\begin{algorithm}[!t]
\caption{The Proposed Algorithm for Solving~\eqref{eq:minv}}
\begin{algorithmic}[1]
\label{alg:Opt2}
\STATE \textbf{Initialization}: $\varrho$, $ \varsigma>1$, $ v>0$, $n=1, q^{(0)}=0, q^{(1)}=1, \mathbf{v}^{(0)}, \overline{\mathbf{v}}^{(0)} \in \widehat{\mathcal{C}}, \alpha_{\overline{\mathbf{v}}}>0, \alpha_{\mathbf{v}}>0, \tilde{\mathbf{v}}^{(1)}=\mathbf{v}^{(1)}=\mathbf{v}^{(0)}, \zeta \in[0,1), b^{(1)}=1, c^{(1)}=f\left(\mathbf{v}^{(1)}\right)$
\STATE \textbf{repeat} \textit{(outer loop: penalty method)}
\STATE \textbf{repeat}  \textit{(inner loop: APG method)}
\STATE \text { Update } $\overline{\mathbf{v}}^{(n)}$ \text { as }
$\overline{\mathbf{v}}^{(n)}\!\!=\!\!\mathbf{v}^{(n)}\!+\!\frac{q^{(n\!-\!1)}}{q^{(n)}}(\tilde{\mathbf{v}}^{(n)}\!\!-\!\!\mathbf{v}^{(n)})\!+\!\frac{q^{(n\!-\!1)}\!\!-\!\!1}{q^{(n)}}(\mathbf{v}^{(n)}\!\!-\!\!\mathbf{v}^{(n\!-\!1)}),$
\STATE Set $\tilde{\mathbf{v}}^{(n+1)}=\mathcal{P}_{\hat{\mathcal{C}}}\left(\overline{\mathbf{v}}^{(n)}-\alpha_{\mathbf{v}} \nabla f\left(\overline{\mathbf{v}}^{(n)}\right)\right)$,
\STATE \textbf{ if } $f\left(\tilde{\mathbf{v}}^{(n+1)}\right) \leq c^{(n)}-\zeta\left\|\tilde{\mathbf{v}}^{(n+1)}-\overline{\mathbf{v}}^{(n)}\right\|^2$ \textbf{then}
\STATE $\mathbf{v}^{(n+1)}=\tilde{\mathbf{v}}^{(n+1)}$
\STATE \textbf{else}
\STATE \text { Update } $\hat{\mathbf{v}}^{(n+1)}$ \text { as \eqref{eq:vhatn+1} } and $\mathbf{v}^{(n+1)}$ \text { as \eqref{eq:vn+1} }
\STATE \textbf{end if}
\STATE Set $q^{(n+1)}=\frac{1+\sqrt{4\left(q^{(n)}\right)^2+1}}{2}$.
\STATE Update $b^{(n+1)}$ \text { as \eqref{eq:bn+1}, and } $c^{(n+1)}$ \text { as \eqref{eq:cn+1} }
\STATE  Set $n=n+1$
\STATE \textbf{until}$\left|\frac{f\left(\mathbf{v}^{(n)}\right)-f\left(\mathbf{v}^{(n-10)}\right)}{f\left(\mathbf{v}^{(n)}\right)}\right|\!\leq\! \epsilon$ \text{or} $\left|\frac{h\left(\boldsymbol{\theta}^{(n)}\right)-h\left(\boldsymbol{\theta}^{(n-1)}\right)}{h\left(\boldsymbol{\theta}^{(n)}\right)}\right|\!\leq\! \epsilon$
\STATE \text { Increase } $\varrho=\varrho \times \varsigma$
\STATE \textbf{until}{ Convergence.}
\end{algorithmic}
\end{algorithm}

 Therefore, for the given penalty coefficients $\mu_1$, $\mu_2$, and $\mu_3$, the penalized objective function of Problem\eqref{eq:minRE}, denoted by $\mathbf{v}$, can be expressed as
\vspace{-0.15em}
\begin{align}\label{eq:minv}
f(\mathbf{v})  \triangleq &\log_{2}\left(1+ \frac{U_E(\boldsymbol\theta)}{V_E(\boldsymbol\theta)}\right)+\varrho[\mu_1 \Psi_{1}(\boldsymbol{\theta})+\mu_2 \Psi_{2}(\mathbf{z})\nonumber\\
&+\mu_3 \Psi_{3}(\boldsymbol{\theta}, \mathbf{z})],   
\end{align}
with $\varrho>0$. Accordingly, at each iteration of the iterative process, the  regularized optimization problem 
\begin{align}\label{eq:minv}
&\min _{\mathbf{v} \in \hat{\mathcal{C}}} f(\mathbf{v}),   
\end{align}
for a given $\varrho$ is solved where  $\widehat{\mathcal{C}}$ represents the convex feasible set of \eqref{eq:minv} (i.e., ${\eqref{eq:st0},\eqref{eq:st1},\eqref{eq:zlk}})$ and  $\mathbf{v} \triangleq\left[\boldsymbol{\theta}^T, \mathbf{z}^T\right]^T$.
We summerize our proposed method  to solve problem \eqref{eq:minv}, which combines the penalty method and the APG method, in Algorithm~\ref{alg:Opt2}. We note that $\mathcal{P}_{C}(\qx)$ in Algorithm~\ref{alg:Opt2} shows the Euclidean projection operator which is defined as
\vspace{-0.1em}
\begin{align}
\mathcal{P}_\mathcal{C}(\qx)=\text{argmin}_{\qu \in \mathcal{C}} \Vert \qx-\qu \Vert. 
\end{align}
It is clear that the two main operations in the implementation
 of Algorithm~\ref{alg:Opt2} are the computation of the gradient 
of the objective function and the projections. 

\emph{1) Gradient  of the objective function:}
The gradients $\frac{\partial}{\partial \theta_{l,k}} f(\mathbf{v})$ and $\frac{\partial}{\partial z_{l,k}} f(\mathbf{v})$ can be calculated as
\begin{align} \label{eq:partialthetafv}
& \frac{\partial}{\partial \theta_{l,k}} f(\mathbf{v})= \frac{\partial}{\partial \theta_{l,k}} \mathrm{R}_E(\mathbf{v})+\varrho \frac{\partial}{\partial \theta_{l,k}} \widetilde{\Psi}(\mathbf{v}), \\
& \label{eq:partialzfv}\frac{\partial}{\partial z_{l,k}} f(\mathbf{v})=\frac{\partial}{\partial z_{l,k}} \mathrm{R}_E(\mathbf{v})+\varrho \frac{\partial}{\partial z_{l,k}} \widetilde{\Psi}(\mathbf{v}),
\end{align}
where
\begin{align} \label{eq:partialthetaRE}
 &\frac{\partial}{\partial \theta_{l,k}} \mathrm{R}_E(\mathbf{v})\notag\\
&=\frac{1}{\log 2}\left[\frac{\frac{\partial}{\partial \theta_{l,k}}\left(U_E(\mathbf{v})+V_E(\mathbf{v})\right)}{\left(U_E(\mathbf{v})+V_E(\mathbf{v})\right)}-\frac{\frac{\partial}{\partial \theta_{l,k}} V_E(\mathbf{v})}{V_E(\mathbf{v})}\right],
\end{align}
with
\vspace{0.2em}
\begin{subequations}
\begin{align}
&\frac{\partial}{\partial \theta_{l,k}} U_E(\mathbf{v})\!=\! 
\begin{cases}
2(\sum\limits_{l \in \mathcal{L}} \!\!\sqrt{\rho_{d}\left(M\!\!-\!\!\left|\mathcal{S}_l\right|\right) \gamma_{l,E}} \theta_{l,1})
\\
\times\sqrt{\rho_{d}\left(M\!-\!\left|\mathcal{S}_l\right|\right) \gamma_{l,E}}\\+ 2 \rho_{d}\theta_{l, 1}(\beta_{l,E}\!-\!\delta_{l,1}\gamma_{l,E}), & k=1, \\
0, & k \neq 1, 
\end{cases} \\
&\frac{\partial}{\partial \theta_{l,k}} V_E(\mathbf{v})= 
\begin{cases}
\!2 \rho_{d} \theta_{l,t}(\beta_{l,E}-\delta_{l,1} \gamma_{l,E}), & t = k \backslash\{1\}, \\
0, & t \neq k.
\end{cases}
\end{align}  
\end{subequations}
\vspace{-0.55em}
In addition,
$\frac{\partial}{\partial \theta_{l,k}} \mathrm{R}_i(\mathbf{v})$ can be calculated as 
\vspace{0.35em}
\begin{align} \label{eq:partialthetaRi}
 &\frac{\partial}{\partial \theta_{l,k}} \!\mathrm{R}_i(\mathbf{v})&\!\!\!\!=\!\!\frac{1}{\log 2}\!\left[\frac{\frac{\partial}{\partial \theta_{l,k}}\!\left(U_i(\mathbf{v})\!+\!V_i(\mathbf{v})\right)}{\left(U_i(\mathbf{v})\!+\!V_i(\mathbf{v})\right)}\!-\!\frac{\frac{\partial}{\partial \theta_{l,k}} V_i(\mathbf{v})}{V_i(\mathbf{v})}\right],
\end{align}
\vspace{-0.3em}
where $\frac{\partial}{\partial \theta_{l,k}} U_i(\mathbf{v})$ and $\frac{\partial}{\partial \theta_{l,k}} V_i(\mathbf{v})$ are given by


\vspace{-0.1em}
\begin{align}
\frac{\partial}{\partial \theta_{l,k}} U_i(\mathbf{v})\!=\! 
\begin{cases}
2(\sum\limits_{l \in \mathcal{L}} \!\!\sqrt{\rho_{d}\left(M\!\!-\!\!\left|\mathcal{S}_l\right|\right) \gamma_{l,k}} \theta_{l,k})\\\times\sqrt{\rho_{d}\left(M\!-\!\left|\mathcal{S}_l\right|\right) \gamma_{l,k}}, & i = k \backslash\{1\}, \\
0, & i \neq k,
\end{cases} 
\end{align}

\begin{align}
\frac{\partial}{\partial \theta_{l,k}} V_i(\mathbf{v})= \begin{cases}\!2 \rho_{d} \theta_{l,k}\left(\beta_{l,k}-\delta_{l,k} \gamma_{l,k}\right), & i = k \backslash\{1\}, \\
\!2 \rho_{d} \theta_{l,k}\left(\beta_{l,i}-\delta_{l,i} \gamma_{l,i}\right), & i \neq k.\end{cases}
\end{align}
Thus, we obtain
\begin{align} \label{eq:Psitelda(v)}
& \frac{\partial}{\partial \theta_{l,k}} \widetilde{\Psi}(\mathbf{v})=4\mu_3  \max \left(0, \theta_{l,k}^2-z_{l k}^2\right) \theta_{l,k} \notag\\
& -\mu_2 \sum_{i \in \mathcal{K}} 2 \max \left(0, \mathrm{R}_{\mathrm{th},k}-\mathrm{R}_i(\boldsymbol{\theta})\right) \frac{\partial}{\partial \theta_{l,k}} \mathrm{R}_i(\mathbf{v}),
\end{align}
\vspace{-0.35em}
\begin{align}
 \frac{\partial}{\partial z_{l,k}} \widetilde{\Psi}(\mathbf{v})=&\mu_1\left(2 z_{l,k}-4 z_{l,k}^3\right)\!-\!4\mu_3 \max \left(0, \theta_{l,k}^2-z_{l,k}^2\right) z_{l,k}\notag \\
& -4 \mu_3 \max \left(0,1-\sum_{l \in \mathcal{L}} z_{l,k}^2\right) z_{l,k}.
\end{align}
\vspace{-0.25em}
\emph{2) Projection onto feasible set $\widehat{\mathcal{C}}$:}
The projection onto the feasible set $\widehat{\mathcal{C}}$ in  Algorithm~\ref{alg:Opt2} can be done by solving the  problem 
\vspace{-0.35em}
\begin{align} \label{eq:PH^}
&\mathcal{P}_{\widehat{\mathcal{C}}}(\mathbf{v}):\min _{\mathbf{v} \in \mathbb{R}^{2LK \times 1}}\|\mathbf{v}-\mathbf{r}\|^2 \\
&\text { s.t. } \quad\eqref{eq:st0},\eqref{eq:st1},\eqref{eq:zlk},\notag
\end{align}
where $\mathbf{r}=\left[\mathbf{r}_1^T, \mathbf{r}_2^T\right]^T \in \mathbb{R}^{2LK \times 1}$ with $\mathbf{r}_1 \triangleq\left[\mathbf{r}_{1,1}^T, \ldots, \mathbf{r}_{1, L}^T\right]^T$ and $\mathbf{r}_{1, l} \triangleq\left[r_{1, l1}, \ldots, r_{1, l K}\right]^T$. Problem \eqref{eq:PH^} can be split into two distinct subproblems as
\vspace{-0.35em}
\begin{align} \label{eq:minthetal}
&\min _{\boldsymbol{\theta}_l \in \mathbb{R}^{LK \times 1}} \left\|\boldsymbol{\theta}_l-\mathbf{r}_{1, l}\right\|^2 \\
&\text { s.t. } \left\|\boldsymbol{\theta}_l\right\|^2 \leq 1, \boldsymbol{\theta}_l \geq 0,\notag
\end{align}
and
\vspace{-0.45em}
 \begin{align} \label{eq:minzl}
&\min _{\mathbf{z}_l \in \mathbb{R}^{LK \times 1}} \left\|\mathbf{z}_l-\mathbf{r}_{2,l}\right\|^2 \\
&\text { s.t. } \mathbf{z}_l \geq 0, \mathbf{z}_l \leq 1,\notag
\end{align}
where the constraints in problems \eqref{eq:minthetal} and \eqref{eq:minzl} adhere to the conditions outlined in \eqref{eq:st0}, \eqref{eq:st1}, and \eqref{eq:zlk}. 
 Solving problem \eqref{eq:minthetal}  involves projecting a given point onto the intersection of a Euclidean ball and the positive orthant, and this projection can be calculated using a closed-form expression  \cite{Hao:2023:EUSIPCO} as 
\begin{align} \label{eq:thetal}
\boldsymbol{\theta}_l=\frac{1}{\max \left(1,\left\|\left[\mathbf{r}_{1, l}\right]_{0}^+\right\|\right)}\left[\mathbf{r}_{1, l}\right]_{0}^+,
\end{align}
where $[\mathbf{x}]_{0}^+\! \triangleq\!\left[\max \left(0, x_1\right), \ldots, \max \left(0, x_K\right)\right]^T, \forall \mathbf{x} \in \mathbb{R}^{K \times 1}$.

We note that in Algorithm~\ref{alg:Opt2}, starting from $\overline{\mathbf{v}}^{(n)}$, we move along the gradient of the objective function using a specific step size
$\alpha_{\overline{\mathbf{v}}}$. By projecting the resulting point $\left(\overline{\mathbf{v}}-\alpha_{\overline{\mathbf{v}}} \nabla f(\overline{\mathbf{v}})\right)$ onto the feasible set $\widehat{\mathcal{C}}$, we obtain
\begin{align} \label{eq:vteldan+1}
\tilde{\mathbf{v}}^{(n+1)}=\mathcal{P}_{\hat{\mathcal{C}}}\left(\overline{\mathbf{v}}^{(n)}-\alpha_{\mathbf{v}} \nabla f\left(\overline{\mathbf{v}}^{(n)}\right)\right).
\end{align}
 However, $f\left(\tilde{\mathbf{v}}^{(n+1)}\right)$ may not improve the objective sequence because $f(\mathbf{v})$ is not convex,thus, $\mathbf{v}^{(n+1)}=\tilde{\mathbf{v}}^{(n+1)}$ is accepted if and only if the objective value $f\left(\tilde{\mathbf{v}}^{(n+1)}\right)$ is below $c^{(n)}$, which represents a relaxation of $f\left(\mathbf{v}^{(n)}\right)$, but relatively close to $f\left(\mathbf{v}^{(n)}\right)$. Following \cite{Li:2015:NIPS}, $c^{(n)}$ is computed as follows:
\vspace{-0.35em}
\begin{align} \label{eq:cn}
    & c^{(n)}= \frac{\sum\limits_{n=1}^\kappa \zeta^{(\kappa-n)} f\left(\mathbf{v}^{(n)}\right)}{\sum\limits_{n=1}^\kappa \zeta^{(\kappa-n)}},
\end{align}
where $\zeta \in[0,1)$ used to control the non-monotonicity degree. After each iteration, $c^{(n)}$ can be iteratively updated as follows:
\begin{align}\label{eq:cn+1}
&{c}^{(n+1)}=\frac{\zeta b^{(n)} c^{(n)}+f\left(\mathbf{v}^{(n)}\right)}{b^{(n+1)}},
\end{align}
where $c^{(1)}\!\!=\!\!f\left(\mathbf{v}^{(1)}\right)$ and $b^{(1)}\!\!=\!\!1$, and ${b}^{(n+1)}$ is calculated as
\vspace{-0.15em}
\begin{align}\label{eq:bn+1}
&{b}^{(n+1)}=\zeta b^{(n)}+1. 
\end{align}
In case the condition $f\left(\tilde{\mathbf{v}}^{(n+1)}\right) \leq c^{(n)}-\zeta\Vert\tilde{\mathbf{v}}^{(n+1)}-\overline{\mathbf{v}}^{(n)}\Vert^2$ is not satisfied, extra correction steps are employed to avoid this situation, where $\|\mathbf{x}\|$ denotes the Euclidean norm of vector $\mathbf{x}$. Specifically, an additional point is calculated with a dedicated step size $\alpha_{\mathbf{v}}$ as 
\begin{align}\label{eq:vhatn+1}
\hat{\mathbf{v}}^{(n+1)}=\mathcal{P}_{\hat{\mathcal{C}}}\left(\mathbf{v}^{(n)}-\alpha_{\mathbf{v}} \nabla f\left(\mathbf{v}^{(n)}\right)\right).
\end{align}
Then, $\mathbf{v}^{(n+1)}$ is updated by comparing the objective values at $\tilde{\mathbf{v}}^{(n+1)}$ and $\hat{\mathbf{v}}^{(n+1)}$ as
\vspace{0.15em}
\begin{align}\label{eq:vn+1}
\mathbf{v}^{(n+1)} \triangleq\left\{\begin{array}{ll}
\tilde{\mathbf{v}}^{(n+1)}, & \text { if } f\left(\tilde{\mathbf{v}}^{(n+1)}\right) \leq f\left(\hat{\mathbf{v}}^{(n+1)}\right),  \\
\hat{\mathbf{v}}^{(n+1)}, & \text {otherwise}.
\end{array}\right.
\end{align}
Since the feasible set $\widehat{\mathcal{C}}$ has bounds, it is valid to assert that $\nabla f(\mathbf{v})$ is Lipschitz continuous, with a known constant value of $J$, i.e.,
\vspace{-0.25em}
\begin{align}\label{eq:deltafx}
\|\nabla f(\mathbf{x})-\nabla f(\mathbf{y})\| \leq J\|\mathbf{x}-\mathbf{y}\|, \forall \mathbf{x}, \mathbf{y} \in \widehat{\mathcal{C}}.
\end{align}
 

\begin{figure}[!t]
			\centering 
			\includegraphics[width=0.42\textwidth]{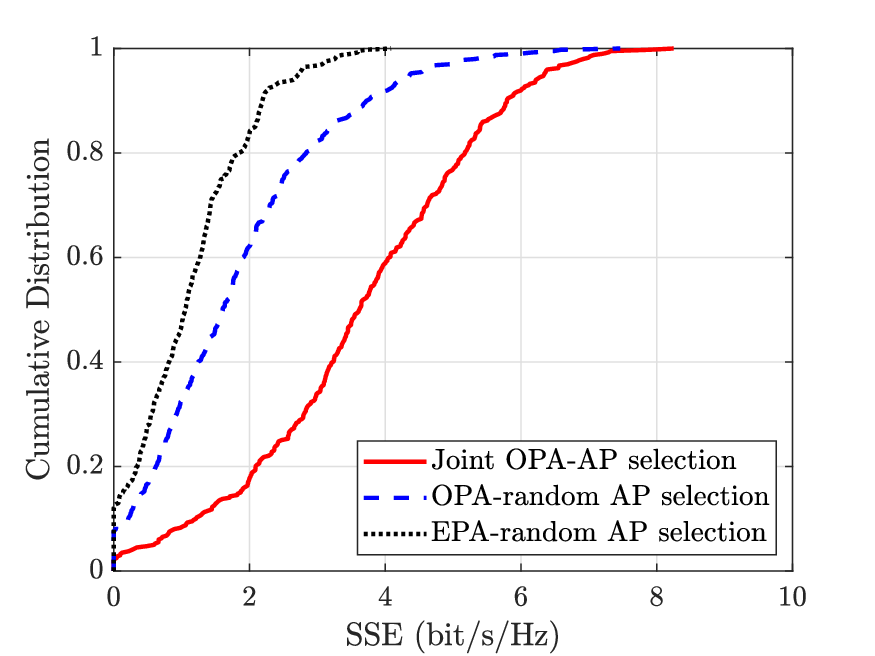}
			\caption{CDFs of the SSE for three cases: EPA-random AP selection, OPA-random AP selection, and joint OPA-AP selection. Here, $L=300$, $M=4$, $K=40$, and $r_E =100$ m.}
			\label{fig1}
   \vspace{1.0em}
\end{figure}

\vspace{-0.7em}
\section{Numerical Results}\label{Section:Simulation}
In this section, we present numerical results to investigate the security performance of   CF-mMIMO using the proposed joint power control and AP selection approach. We assume that all $L$ APs and $K$ users are randomly distributed within a square of $1 \times 1$ km$^2$. To avoid the boundary effects, we use a wrapped-around technique.  Furthermore, Eve is randomly located inside a circle of radius $r_E$ around the attacked user. The large-scale fading $\beta_{l,k}$ is formulated as follows:
\vspace{-0.35em}
\begin{align}
\beta_{l,k} = 10^{\frac{\text{PL}_{l,k}^d}{10}}10^{\frac{F_{l,k}}{10}},
\end{align}
where $10^{\frac{\text{PL}_{l,k}^d}{10}}$ is the path loss, and $10^{\frac{F_{l,k}}{10}}$ denotes the shadowing effect with $F_{l,k}\in\mathcal{N}(0,4^2)$ (in dB).  Besides, $\text{PL}_{l,k}^d$ (in dB) can be calculated as
\begin{align}
\label{PL:model}
\text{PL}_{l,k}^d = -30.5-36.7\log_{10}\Big(\frac{d_{l,k}}{1\,\text{m}}\Big),
\end{align}
and the correlation among the shadowing terms from the $l$-th AP to different $k$ users can be given by  
\vspace{0.55em}
\begin{align}
	 \mathbb{E}\{F_{l,k}F_{j,k'}\} \triangleq \begin{cases} 4^22^{-\vartheta_{k,k'}/9\,\text{m}}, & j = l, \\ 0, & \text{otherwise}, \end{cases}
\end{align}
where $\vartheta_{k,k'}$ is the physical distance between users $k$ and $k'$ \cite{Emil:TWC:2020}. 
In addition, we choose the bandwidth $B=20$ MHz,  a noise power equal to  $-92$ dBm, and the maximum
transmit power for each AP and each user  as $1$ W  and $100$ mW, respectively. Also, $\mathrm{R}_{\mathrm{th},k}=0.2, \forall k$, and $\tau_{p}=K$. 
In Fig. \ref{fig1}, we investigate the effectiveness of using joint power allocation and AP selection approach provided in Algorithm \ref{alg:Opt2}. The cumulative distribution function of the SSE is shown for three cases: equal power allocation (EPA) with random AP selection, optimum power allocation (OPA) and random AP selection, and our proposed APG-based joint power allocation and AP selection scheme.
We can see that optimum power control enhances the  $50\%$ likely SSE  performance up to $65\%$ compared with equal power control. Also, it is observed that the proposed joint optimization approach can provide a significant improvement in the SSE with a gain of up to $265\%$ and $120\% $ compared with the cases relying on random AP selection with equal power control and optimum power control, respectively. These results highlight the effectiveness of our joint
optimization approach over heuristic schemes.

\begin{figure}[!t]
   \vspace{-0.5em}
			\centering 
			\includegraphics[width=0.42\textwidth]{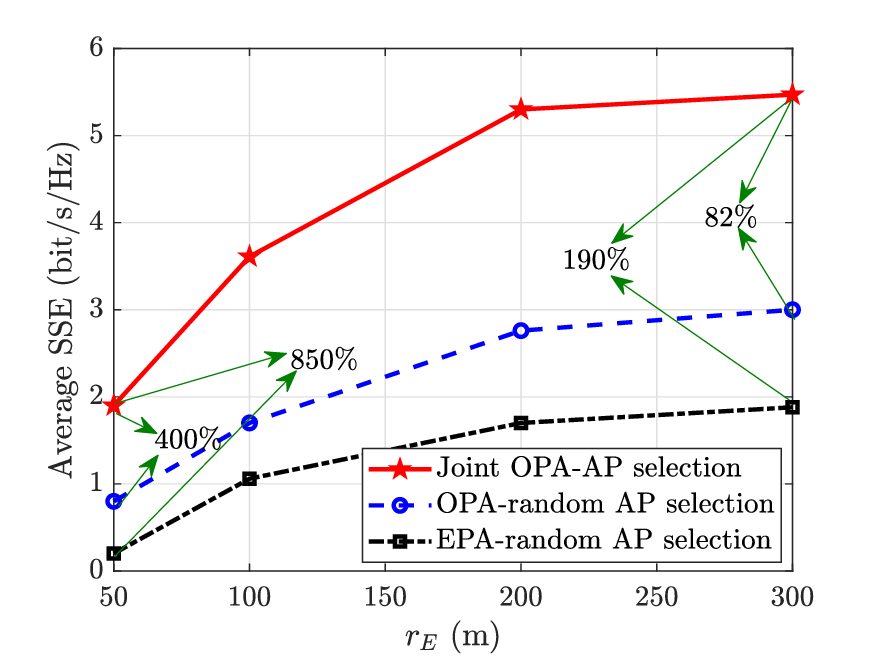}
			\caption{Average SSE  vs $r_E$, $L=300$, $M=4$, and $K=40$.}
			\label{fig2}
   \vspace{0.9em}
\end{figure}

In Fig. \ref{fig2}, we investigate the impact of Eve's position on the SSE performance. In particular, we plot the average SSE, while the average is taken over the large-scale fading realizations, versus the radius of a circle surrounding user $1$, $r_E$, as Eve's potential positions. We can see that SSE remarkably improves when Eve moves away from target user {$1$}. More importantly, the performance gap between the joint optimization approach and the heuristic ones is more pronounced for more challenging scenarios when Eve is located in the closer vicinity of the target user. Specifically, when $r_E$ changes from $200$ m to $50$ m, the average SSE performance gap between the joint optimization approach and EPA with random AP selection benchmark increases from $190\%$ to $850\%$.


\section {Conclusion}
This paper investigated the SSE performance of a CF-mMIMO system in the presence of active eavesdropping, employing PPZF precoding with imperfect channel estimation. 
 To this end, 
 we proposed a joint optimization approach to minimize the   SE at  Eve under a transmit
power constraint at the APs and  QoS SE requirements at the users. 
We showed that our
jointly optimized APG-based approach provides substantially higher SSE compared to the heuristic approaches, especially in the scenarios where Eve is located in the close vicinity of the targeted user.


\vspace{0em}
\balance
\bibliographystyle{IEEEtran}
\bibliography{bibliography.bib}

\begin{thebibliography}{10}
\providecommand{\url}[1]{#1}
\csname url@samestyle\endcsname
\providecommand{\newblock}{\relax}
\providecommand{\bibinfo}[2]{#2}
\providecommand{\BIBentrySTDinterwordspacing}{\spaceskip=0pt\relax}
\providecommand{\BIBentryALTinterwordstretchfactor}{4}
\providecommand{\BIBentryALTinterwordspacing}{\spaceskip=\fontdimen2\font plus
\BIBentryALTinterwordstretchfactor\fontdimen3\font minus \fontdimen4\font\relax}
\providecommand{\BIBforeignlanguage}[2]{{%
\expandafter\ifx\csname l@#1\endcsname\relax
\typeout{** WARNING: IEEEtran.bst: No hyphenation pattern has been}%
\typeout{** loaded for the language `#1'. Using the pattern for}%
\typeout{** the default language instead.}%
\else
\language=\csname l@#1\endcsname
\fi
#2}}
\providecommand{\BIBdecl}{\relax}
\BIBdecl

\bibitem{Hien:JWCOM:2017}
H.~Q. Ngo, A.~Ashikhmin, H.~Yang, E.~G. Larsson, and T.~L. Marzetta, ``Cell-free massive {MIMO} versus small cells,'' \emph{{IEEE} Trans. Wireless Commun.}, vol.~16, no.~3, pp. 1834--1850, Mar. 2017.

\bibitem{Timilsina:GC:2018}
S.~Timilsina, D.~Kudathanthirige, and G.~Amarasuriya, ``Physical layer security in cell-free massive {MIMO},'' in \emph{Proc. IEEE GLOBECOM}, Dec. 2018.

\bibitem{Hoang:TCOM:2018}
T.~M. Hoang, H.~Q. Ngo, T.~Q. Duong, H.~D. Tuan, and A.~Marshall, ``Cell-free massive {MIMO} networks: Optimal power control against active eavesdropping,'' \emph{{IEEE} Trans. Commun.}, vol.~66, no.~10, pp. 4724--4737, Oct. 2018.

\bibitem{Salah:Globecom:2020}
S.~Elhoushy and W.~Hamouda, ``Nearest {APs}-based downlink pilot transmission for high secrecy rates in cell-free massive {MIMO},'' in \emph{Proc. GLOBECOM}, Dec. 2020.

\bibitem{Alageli:TIFS:2020}
M.~Alageli, A.~Ikhlef, F.~Alsifiany, M.~A.~M. Abdullah, G.~Chen, and J.~Chambers, ``Optimal downlink transmission for cell-free {SWIPT} massive {MIMO} systems with active eavesdropping,'' \emph{{IEEE} Trans. Inf. Forensics Security}, vol.~15, pp. 1983--1998, Nov. 2019.

\bibitem{Zhang:SYS.2020}
X.~Zhang, D.~Guo, K.~An, and B.~Zhang, ``Secure communications over cell-free massive {MIMO} networks with hardware impairments,'' \emph{{IEEE} Syst. J.}, vol.~14, no.~2, pp. 1909--1920, Jun. 2020.

\bibitem{Zhang:TVT:2010}
X.~Zhang, T.~Liang, K.~An, G.~Zheng, and S.~Chatzinotas, ``Secure transmission in cell-free massive {MIMO} with {RF} impairments and low-resolution {ADCs/DACs},'' \emph{{IEEE} Trans. Veh. Technol.}, vol.~70, no.~9, pp. 8937--8949, Sep. 2021.

\bibitem{Hien:Asilomar:2018}
H.~Q. Ngo, H.~Tataria, M.~Matthaiou, S.~Jin, and E.~G. Larsson, ``On the performance of cell-free massive {MIMO} in {Ricean} fading,'' in \emph{Proc. IEEE Asilomar Conf. Signals, Systems, and Computers}, Nov. 2018.

\bibitem{Ammar:GLOBSIP:2019}
H.~A. Ammar and R.~Adve, ``Power delay profile in coordinated distributed networks: {User}-centric v/s disjoint clustering,'' in \emph{Proc. IEEE GlobalSIP}, Nov. 2019.

\bibitem{Buzzi:TWC:2020}
S.~Buzzi, C.~D’Andrea, A.~Zappone, and C.~D’Elia, ``User-centric {5G} cellular networks: {Resource} allocation and comparison with the cell-free massive {MIMO} approach,'' \emph{{IEEE} Trans. Wireless Commun.}, vol.~19, no.~2, pp. 1250--1264, Feb. 2020.

\bibitem{Chen:JSAC:2021}
S.~Chen, J.~Zhang, E.~Björnson, J.~Zhang, and B.~Ai, ``Structured massive access for scalable cell-free massive {MIMO} systems,'' \emph{{IEEE} J. Sel. Areas Commun.}, vol.~39, no.~4, pp. 1086--1100, Apr. 2021.

\bibitem{Hao:2023:EUSIPCO}
C.~Hao, T.~Vu, H.-Q. Ngo, M.~Dao, X.~Dang, and M.~Matthaiou, ``User association and power control in cell-free massive {MIMO} with the {APG} method,'' in \emph{Proc. IEEE EUSIPCO}, Sep. 2023.

\bibitem{Yasseen:TWC:2023}
Y.~S. Atiya, Z.~Mobini, H.~Q. Ngo, and M.~Matthaiou, ``Secure transmission in cell-free massive {MIMO} under active eavesdropping,'' \emph{IEEE Trans. Wireless Commun}, 2023, submitted.

\bibitem{Emil:TWC:2020}
G.~Interdonato, M.~Karlsson, E.~Björnson, and E.~G. Larsson, ``Local partial zero-forcing precoding for cell-free massive {MIMO},'' \emph{{IEEE} Trans. Wireless Commun.}, vol.~19, no.~7, pp. 4758--4774, Jul. 2020.

\bibitem{YASSEEN:2023:VTC}
Y.~S. Atiya, Z.~Mobini, H.~Q. Ngo, and M.~Matthaiou, ``Cell-free massive {MIMO} with protective partial zero-forcing and active eavesdropping,'' in \emph{Proc. IEEE VTC}, Jun. 2023.

\bibitem{Vu:IOT:2022}
T.~T. Vu, T.~N. Duy, H.~Q. Ngo, M.~N. Dao, N.~H. Tran, and R.~H. Middleton, ``Joint resource allocation to minimize execution time of federated learning in cell-free massive {MIMO},'' \emph{{IEEE} Internet Things J.}, vol.~9, no.~21, pp. 21\,736--21\,750, Nov. 2022.

\bibitem{Mai:TWC:2022}
T.~C. Mai, H.~Q. Ngo, and L.-N. Tran, ``Energy efficiency maximization in large-scale cell-free massive {MIMO}: A projected gradient approach,'' \emph{{IEEE} Trans. Wireless Commun.}, vol.~21, no.~8, pp. 6357--6371, Aug. 2022.

\bibitem{farooq2021utility}
M.~Farooq, H.~Q. Ngo, E.-K. Hong, and L.-N. Tran, ``Utility maximization for large-scale cell-free massive {MIMO} downlink,'' \emph{IEEE Trans. Commun.}, vol.~69, no.~10, pp. 7050--7062, Nov. 2021.

\bibitem{Li:2015:NIPS}
H.~Li and Z.~Lin, ``Accelerated proximal gradient methods for nonconvex programming,'' in \emph{Proc. NeurIPS}, vol.~1, no.~9, Dec. 2015.

\end{thebibliography}
\end{document}